\documentclass[12pt,showpacs,preprintnumbers]{revtex4}

\usepackage{graphicx}
\usepackage{dcolumn}
\usepackage{bm}
\usepackage{amssymb}
\usepackage{epsfig}    
\def\beq{\begin{equation}}
\def\eeq{\end{equation}}
\def\ba{\begin{array}}
\def\ea{\end{array}}
\def\bea{\begin{eqnarray}}
\def\eea{\end{eqnarray}}

\def\sq2{\sqrt{2}}
\def\End{\end{document}}

\def\NPB{{\em Nucl. Phys.} B}
\def\PLB{{\em Phys. Lett.}  B}
\def\PRL{\em Phys. Rev. Lett.}
\def\PRD{{\em Phys. Rev.} D}

\def\EPC{{\em Euro. Phys. J.} C}

\begin{document}                                                              

\title{QCD correction to single top quark production at
the ILC}%
\author{%
{F.~Pe\~nu\~nuri\,$^1$},~~{F.~Larios\,$^2$}~~and~~
{Antonio~O.~Bouzas\,$^2$}  }
\affiliation{%
\vspace*{2mm} 
$^1$Facultad de  Ingenieria, Universidad Aut\'onoma de
Yucat\'an, A.P. 150, Cordemex, M\'erida, Yucatan, M\'exico. \\
$^2$Departamento de F\'{\i}sica Aplicada,
CINVESTAV-M\'erida, A.P. 73, 97310 M\'erida, Yucat\'an, M\'exico
}

\begin{abstract}
\hspace*{-0.35cm}

Single top quark production at the ILC can be used to obtain a
high precision measurements of the the $V_{tb}$ CKM matrix
element as well as the effective $tbW$ coupling.
We have calculated the QCD correction for the cross section in
the context of an effective vector boson approximation.
Our results show a $\sim 10\%$ increase due to the strong
interaction.


\end{abstract}

\maketitle

\setcounter{footnote}{0}
\renewcommand{\thefootnote}{\arabic{footnote}}

\section{Introduction}

The top quark stands out as the heaviest known elementary
particle and its properties and interactions are one of the
most important measurements for present and future high
energy colliders\cite{topreviews}.  At the Tevatron and
at the LHC the process of single Top quark production
has been extensively studied \cite{singlet}.

The top quark is likely to provide us with the first clues
of physics beyond the Standard Model \cite{newphys}.
In fact, new physics effects are probably already manifest
in the recent forward-backward asymmetry observed at the
Tevatron \cite{fbteva, fbasy}.

The planned International Linear Collider (ILC) will collide
electron and positron beams at an initial energy of 500 GeV
and higher.  It will provide a clean environment for the
study of precision measurements.

The single top production processes at lepton and photon
($e^+e^-$, $e^- e^-$, $\gamma e$ and $\gamma \gamma$)
colliders have been extensively studied at tree level in
Ref.~\cite{boos}.  The reaction $\gamma e^- \to \bar t b \nu_e$,
is particularly suitable for precision studies, as it does not
have the $t\bar t$ background.  Compared to the ILC
$e^+ e^- \to t\bar b e^- \bar \nu_e$ process the $\gamma e^-$
reaction can yield a larger production rate and is directly
proportional to the $V_{tb}$ term.  Further studies,  have thus
been done for this reaction. In particular,
the QCD corrections have been studied in Ref.~\cite{kuhn}.
Their conclusion is that the QCD correction is not very large
($\sim 5\%$) so that this mode remains very well suited for a
precise measurement of $V_{tb}$.  
The approach used by \cite{kuhn} is
to use the effective vector boson approximation, also known as 
{\it effective W-approximation} \cite{dawson85} (EWA) and to
compute the QCD loop corrections for the $W^+ \gamma \to t\bar b$
fusion process.  Then, the convolution with the $f_{W^+/e^+} (x)$
distribution function is applied to obtain the correction to
the actual $e^+\gamma$ process.  We would like to point out
that the authors in Ref.~\cite{kuhn} have made a very clear
and thorough presentation of the calculation.  In this work we
use their analysis on the $W^+ \gamma \to t\bar b$ process to
estimate the QCD correction for the
$e^+ e^- \to t\bar b e^- \bar \nu_e$ process of the ILC.
Here, in addition to the convolution with the $W^+$ boson
we will use the effective photon (as well as the
effective Z-boson) approximation to obtain the QCD correction.
We will use the same input values for masses and coupling
constants, except for the masses of top and bottom quarks we
take $m_t=173$ GeV and $m_b=4.2$ GeV.


\section{Vector boson contributions at tree level}

\begin{figure}
\includegraphics[width=9.5cm,height=5.5cm]{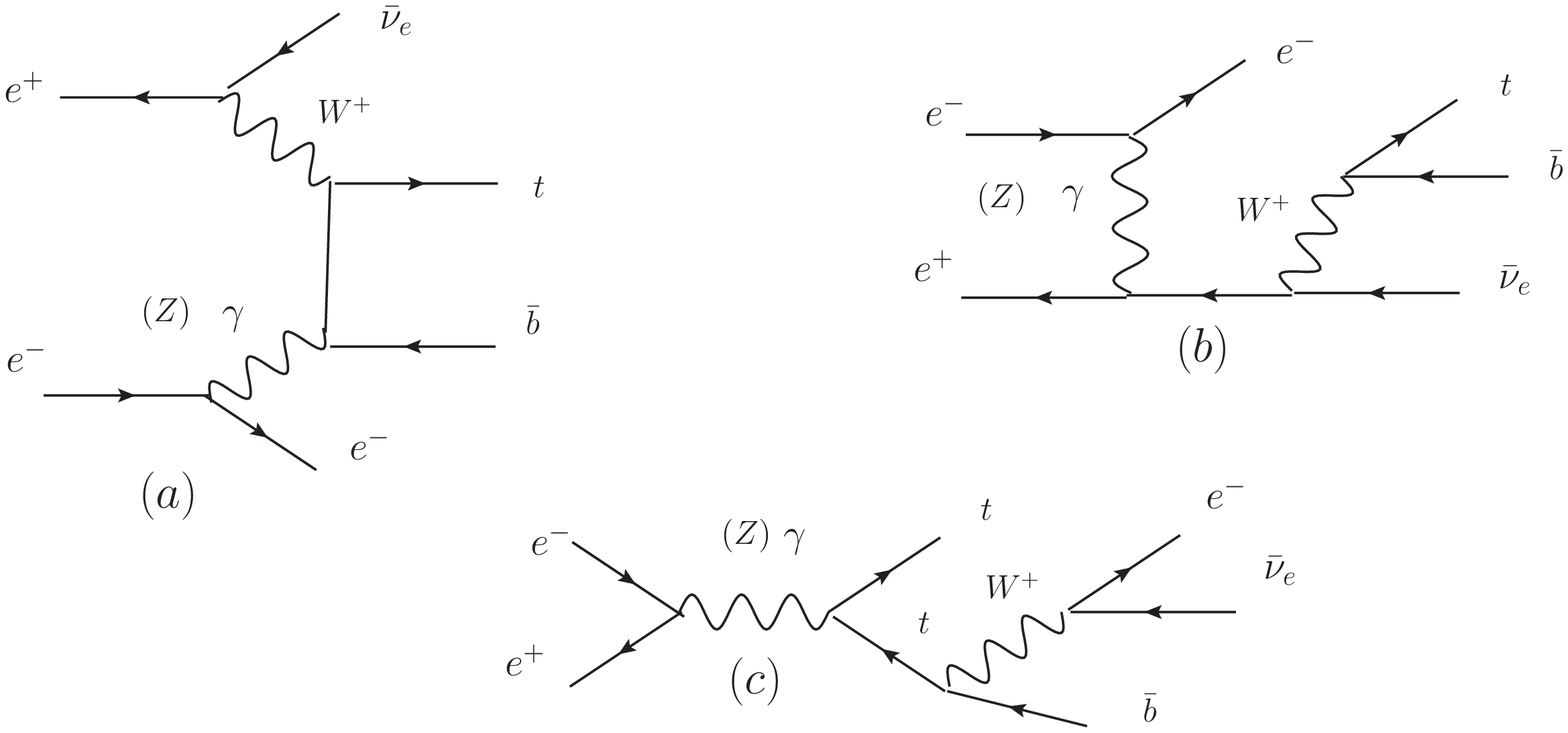}
\caption{The three type of diagrams for the
$e^+e^- \to t\bar b e^- \bar \nu_e$ process.}
\label{diagrams}
\end{figure}

At tree level there are 20 diagrams for the
$e^+ e^- \to t \bar b e^- \bar \nu_e$ process\cite{boos}.
We can list them in three different types: (a) vector boson
fusion, (b) vector boson exchange and (c) $e^+ e^-$ annihilation
(see Figure\ref{diagrams}).
For the energy range we consider one of the diagrams actually
corresponds to $t\bar t$ production, where one of the tops decays
leptonically. 
In order to exclude $t\bar t$ production from the single top
process we discard
all events where the invariant mass of the decay products
($e^-$,$\bar \nu_e$,$\bar b$) falls inside an interval around
the top mass $m_t -\Delta M \leq M_{e\nu b} \leq m_t+\Delta M$.
We take the value $\Delta M =20$GeV as in Ref.~\cite{boos}.

The effective-W approximation relies on the fact that
the vector fusion diagrams become dominant when heavy particles
are produced at very high energy collisions
\cite{dawson85}.  In general, 3 conditions should be met for the
EWA to work well:  (1) The mass of the vector boson
($M_W$ or $M_Z$) should be much smaller than its energy, and this
can be met if we require $M_V \ll \sqrt{s}/2$, (2) for $q\bar q$
production $m_q \gg M_V$, this is true for the top quark but
not for the bottom quark, and (3) One polarization mode should
be dominant so that interference effects can be neglected.
Fortunately, in our case the mode $W \gamma \to t\bar b$
dominates for longitudinal $W$, and the modes with the $Z$
boson $W Z \to t\bar b$ give even lower contributions.

\begin{figure}
\includegraphics[width=9.5cm,height=2.5cm]{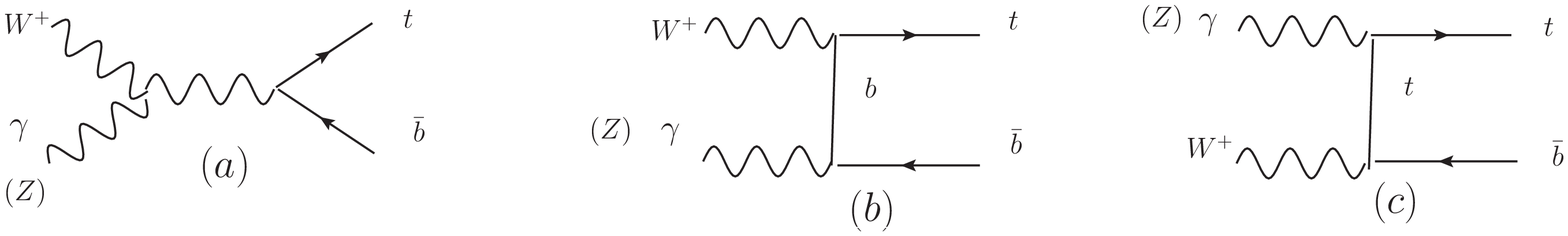}
\caption{The vector boson fusion diagrams for the
$W^+ \gamma (Z) \to t\bar b$ process.}
\label{wgtb}
\end{figure}

\begin{figure}
\includegraphics[width=5.0cm,height=4.5cm]{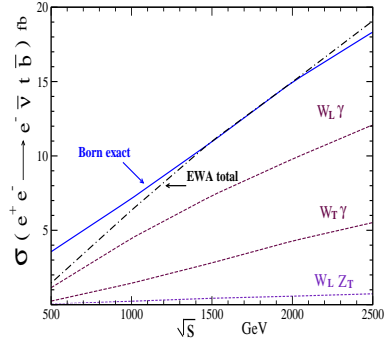}
\caption{The contributions from $W^+\gamma$ and $W^+Z$
fusion to the $e^+e^- \to t\bar b e^- \bar \nu_e$ process.
The solid line shows the exact calculation.}
\label{bornsigma}
\end{figure}

As expected, this method works very well for
$t\bar t$ production at high $\sqrt{s}$ and to a lesser degree
for single top, which in our case can be seen as $t\bar b$
production.  In Ref.~\cite{kuhn} the QCD correction to the
process $e^+ \gamma \to \bar t\bar b \bar \nu_e$ was calculated
by doing first the QCD correction to the $W^+ \gamma$ fusion
into $t\bar b$ and then by taking the convolution with an
effective $W^+$ coming from the initial positron (see Figure
\ref{wgtb}).  We follow the same approach by doing the
one loop QCD correction to $W^+ \gamma \to t\bar b$ as well
as $W^+ Z \to t\bar b$ and then convolute with the effective
distribution functions for $W^+$, $\gamma$ and $Z$:
\bea
 \sigma(e^+ e^- \to t\bar{b}\bar{\nu}_e e^-) &&= 
\label{convolution} \\
&&\sum_{W_L , W_T}  \int_{x_W^{min}}^1 dx_W f_{W^+/e^+}(x_W)
 \int_{0}^1 dx_\gamma f_{\gamma/e^-}(x_\gamma)
 \; \sigma(W^+ \gamma \to t\bar{b})( \hat s )  
\nonumber \\
+ &&\sum_{W_{L,T} , Z_{L,T}}  
\int_{x_W^{min}}^1 dx_W f_{W^+/e^+}(x_W)
\int_{x_Z^{min}}^1 dx_Z f_{Z/e^-}(x_Z)
 \; \sigma(W^+ Z \to t\bar{b})(\hat s)  \nonumber
\eea
Where, $x_V^{min} = 2 M_V/\sqrt{s}$, $\hat s = x_W x_\gamma s$
and the structure functions can be found in \cite{dawson85}.
The tree level cross section for the single top production at
the ILC is shown in Fig.~\ref{bornsigma}.  The exact Born level
calculation for the $e^+ e^- \to t \bar b e^- \bar \nu_e$ process
is obtained with Calchep \cite{pukhov} and is shown by the solid
line.  We can see that the prediction of the EWA (dot-dashed curve)
is in very good agreement with the exact result for center of mass
energies above 1.5 TeV.  However, for the energy range of the ILC
the EWA values can be significantly lower.  In particular, for
$\sqrt{s} = 1000$GeV there is a $15\%$ difference and for
$\sqrt{s} = 500$GeV the EWA result be about one half of the exact
value.

There is one aspect of the calculation that is worth mentioning.
Because of the kinematics of the  $W^+ Z \to t\bar b$  process, we
run into a divergent behavior as we integrate over the Mandelstam
variable $t$ (or the polar angle of the outgoing quark). 
At a certain value of $t$ the massive $Z$ boson can actually
decay into $b\bar b$ and this makes the bottom quark
propagator to hit a pole at this value.  We were able to avoid
this singularity by setting $k^2_Z=0$ instead of $M_Z$.
This is completely justifiable in the context of the EWA.
Let's understand more the importance of the assumption
$M_V \ll \sqrt{s}/2$.  In the complete
process (like $e^+ e^- \to W^{+*} Z^* \to t\bar b e^- \bar \nu_e$)
the virtual $Z$ gets a space-like momenta $k^2_Z \le 0$ and is
always far from the on-shell condition. In fact, the EWA
works better when the initial state vector boson
momentum square is set equal to zero: $k^2_Z =0$, $k^2_W=0$
(see Ref.~\cite{kauffman} for a detailed discussion). 
Nevertheless, when dealing with a process like $t\bar t$
production one may set $k^2_Z = M^2_Z$ as this introduces only
a small error of order $m_Z/\sqrt{s}$.  It is customary to set
the external massive $W^+$ and $Z$ on-shell for convenience.
However, for the single top process
the fact that $Z$ is heavy enough to decay into $b\bar b$ is
prompting us to implement the $k^2_Z=0$ condition in order to
avoid the divergent behavior.   Notice that a similar situation
does not apply to the $W^+$ boson as it cannot decay into
$t\bar b$. Therefore in our study we choose to keep the on-shell
condition $k^2_W=M^2_W$ for the initial state $W^+$ but impose
$k^2_Z=0$ for the $Z$ boson.  For the case of the $W^+$ we have
checked that indeed by setting $k^2_W =0$ we don't find a
significant change in the result.

Below, we will describe the QCD corrections to the $W^+ \gamma$
and $W^+ Z$ processes, including the Dipole substraction method
of infrared divergencies.  We have followed closely the analysis
done for the $W^+ \gamma$ mode done by Kuhn et.al. in
Ref.~\cite{kuhn}.

\section{QCD correction to the $W^+ \gamma (Z)\to t \bar b$ process.}

The QCD loop correction to the $W^+ \gamma (Z)\to t \bar b$ process
is given by 9 Feynman diagrams (see Fig.2 of \cite{kuhn}). 
The renormalization procedure involves only the quark's wave
function and mass parameter. Specific formulas can be found
in \cite{kuhn}.  Concerning the renormalization scale dependence
we have also set $\alpha_s$ at the scale $\mu=\sqrt{s}$ for
our numerical calculation (it becomes $\sqrt{\hat s}$ under
the convolution).
The extraction of IR singularities is done with the substraction
method of the dipole formalism \cite{catani}.  This method consists
of adding and substracting a so-called dipole term:
\bea
\sigma^{NLO}(W^+\gamma \to t\bar b) =
\int_{tbg} \left[ (d\sigma^R)_{\epsilon =0} -
(d\sigma^B \otimes dV_{dipole})_{\epsilon =0} \right]
+\int_{tb} [d\sigma^V + d\sigma^B \otimes {\mathbf I}]_{\epsilon =0}
\label{substraction}
\eea
Where $d\sigma^R$ comes from the real emission
$W^+ \gamma (Z) \to t{\bar b} g$ process and
$d\sigma^B \otimes dV_{dipole}$ is the substracting dipole term
that matches point-wise the singularities associated to the
soft and/or collinear gluon.  Both terms are calculated in
$d=4$ dimensions.  In the second integral the same dipole
term has been partially integrated in the gluon phase space
and then added to the virtual correction $d\sigma^V$.
This sum is performed in $d=4-2\epsilon$ dimensions (consistent
with the dimensional regularization).

The general formula for the dipole term is found in Eq.~(5.16)
of \cite{catani}.  The specific expression in our case is:
\bea
d\sigma^B \otimes dV_{dipole} &&=
\frac{ \langle V_{gt,b} \rangle }{2 k_g\cdot k_t}
|{\cal M}_0(\tilde k_{gt},\tilde k_b)|^2 \; +\; 
\{ t\leftrightarrow b\}\, , \label{dipole}
\eea
where
\bea
 \langle V_{gt,b} \rangle &&= 8\pi \alpha_s C_F \{
\frac{2}{1-\tilde z_t (1-y_{gt,b})} -
\frac{\tilde v_{gt,b}}{v_{gt,b}} [1+\tilde z_t+
\frac{m_t^2}{k_g \cdot k_t}] \} \, ,\nonumber \\
\tilde z_t &&= \frac{k_t\cdot k_b}{(k_t + k_g)\cdot k_b}\, ,
\;\;\;\;\;\; y_{gt,b} \, =\, 2\frac{k_g\cdot k_t}{sx_{tb}}\, ,
\;\;\;\;\;\; \tilde v_{gt,b} \, =\, \frac{\lambda_{tb}}{x_{tb}}\, ,
\nonumber \\
v_{gt,b} &&= \sqrt{(1+a_{gt,b})^2-a_{gt,b}^2/z_b}\, ,\;\;\;\;\;\;\;\;
a_{gt,b} \, =\, \frac{2z_b}{x_{tb} (1-y_{gt,b})}\, , \nonumber \\
{\tilde k_b}&&= \frac{x_b}{2} P+\frac{\lambda_{tb}}{\lambda_{gt}} 
(k_b-\frac{P\cdot k_b}{\sqrt{s}} P)\, , \;\;\;\;\;
{\tilde k_{gt}} = P-{\tilde k_b}\, , \;\;\;\;\; P=k_W+k_\gamma
\, ,\nonumber
\eea
and ${\cal M}_0(\tilde k_{gt},\tilde k_b)$ is the
Born level $W^+ \gamma \to t\bar b$ amplitude with one
modification: the final state momenta $k_t$ and $k_b$ have
been replaced by $\tilde k_{gt}$ and $\tilde k_b$ respectively.

The other variables are defined as in \cite{kuhn}:
$\mu_q = {m_q}/{\sqrt{s}}$, $z_q = \mu_q^2$,
$x_t = 1+z_t-z_b$, $x_b=1+z_b-z_t$, $x_{tb} = 1-z_t-z_b$,
$\lambda_{tb} = \lambda (1,z_t,z_b)$,
$\lambda_{gt} = \lambda (1,{(k_g+k_t)^2}/{s},z_b)$,
and $\lambda (x,y,z) = \sqrt{x^2+y^2+z^2-2xy-2xz-2yz}$.

For the real emission correction we have prepared a Fortran
program that integrates the cross section for the
$W^+ \gamma \to t\bar b g$ process along with dipole
substraction.  As it turns out, the substraction term
defined by the dipole formalism in the first
integral of Eq.~(\ref{substraction}) is actually a very
good approximation to the real emission cross section in
an important part of the $tbg$ phase space, so that the
numerical results we obtained were very small:
about two orders of magnitude below the values obtained
for the virtual correction.

The expression for the dipole term in the virtual correction
is:
\bea
d\sigma^B \otimes {\mathbf I} =
|{\cal M}_d (W^+\gamma \to t\bar b)|^2 
\frac{\alpha_s}{2\pi} \frac{1}{\Gamma (1-\epsilon)} 
\left( \frac{4\pi \mu^2}{s} \right)^\epsilon
\; ({\bf I}_{gt,b} + {\bf I}_{gb,t}) \, ,
\eea
where ${\cal M}_d (W^+\gamma \to t\bar b)$ is the Born level
amplitude in $d=4-2\epsilon$ dimensions (the flux term of the
$t\bar b$ phase space integration is understood).  The dipole
function is given by ${\bf I}_{gt,b} = C_F [2I^{eik}+
I^{coll}_{gt,b}]$
( also ${\bf I}_{gb,t}={\bf I}_{gt,b} \{t\leftrightarrow b\}$),
where $I^{eik}$ and $I^{coll}_{gt,b}$ are given by Eqs.~(5.34)
and (5.35) in \cite{catani}:
\bea
I^{eik} &&= \frac{x_{tb}}{\lambda_{tb}}\; \{ 
\frac{\ln \rho}{2\epsilon} + \frac{\pi^2}{6} 
- \ln{\rho} \ln{[1-(\mu_t+\mu_b)^2]}
-\frac{1}{2} ln^2{\rho_t} -\frac{1}{2} ln^2{\rho_b} 
 \nonumber \\ &&+
2Li_2(-\rho)-2Li_2(1-\rho)
-\frac{1}{2} Li_2(1-\rho^2_t) -\frac{1}{2} Li_2(1-\rho^2_b)
\} \nonumber \\
I^{coll}_{gt,b} &&= \frac{1}{\epsilon} +3+\ln{\mu_t}
+\ln{(1-\mu_b)}-2\ln{[(1-\mu_b)^2-z_t]}-\frac{\mu_b}{1-\mu_b}
\label{coll} \\ 
&&-\frac{2}{x_{tb}} \left[ \mu_b (1-2\mu_b) +
z_t \ln{\frac{\mu_t}{1-\mu_b}} \right] \nonumber
\eea
where $\rho^2 = (x_{tb}-\lambda_{tb})/(x_{tb}+\lambda_{tb})$,
$\rho_t = (x_{tb}-\lambda_{tb}+2z_t)/(x_{tb}+\lambda_{tb}+2z_t)$,
and $\rho_b = \rho_t \{ t\leftrightarrow b \}$.  These
formulas also appear in \cite{kuhn}, except that in their
Eq.~(4.14) $I^{coll}_{gt,b}$ the constant term should not be
$5$ but $3$.

Concerning the calculation of $d\sigma^V$, the details
can be found in Ref.~\cite{kuhn}.  We actually worked out this
same computation before doing the case for the $Z$ boson.
As expected from the results shown in Fig.~\ref{bornsigma}
the contribution from the $W^+ Z$ fusion is much smaller
than the one from $W^+ \gamma$.  In fact, we only considered
the correction for the polarizations $W^+$ longitudinal and
$Z$ transversal as the other possibilities are negligible.

\begin{figure}
\includegraphics[width=5.0cm,height=4.5cm]{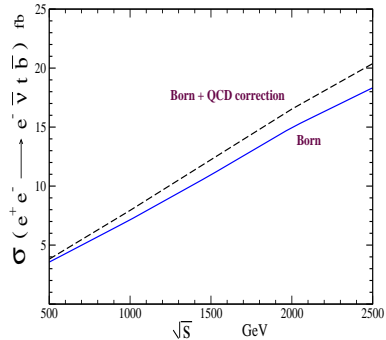}
\caption{The QCD correction from $W^+\gamma$ and $W^+Z$
fusion to the $e^+e^- \to t\bar b e^- \bar \nu_e$ process.
The solid line shows the exact Born level calculation.
The Born plus QCD correction is shown in the dashed line.}
\label{totsigma}
\end{figure}

Our results are shown in Fig.~\ref{totsigma}.  The QCD
correction for the single top production in the $e^+ e^-$
collision process is of order $10\%$ of the Born level
cross section.  It will be interesting to compare this
result based on the effective W-approximation with a
future more robust calculation based on the complete
$e^+ e^- \to t \bar b e^- \bar \nu_e$ process.

\vspace*{4cm}

\noindent
{\bf Acknowledgments}~~~
We thank RedFAE, Conacyt and SNI for support.
F.L. thanks C.-P. Yuan for useful discussions.



\end{document}